\documentclass[twocolumn,letterpaper]{IEEEAerospaceCLS}

\usepackage{amsmath,amssymb,mathtools}
\usepackage{array}
\usepackage{booktabs}
\usepackage{float}
\usepackage{graphicx}
\usepackage{booktabs}
\usepackage{stfloats}
\usepackage{siunitx}
\usepackage{cleveref}
\usepackage[hyphens]{url}

\graphicspath{{figures/}}
\newcommand{\DDeployments}{96}

\newcommand{\DAttempts}{2773}

\newcommand{\DMpcProb}{0.902}
\newcommand{\DMpcTime}{78.3}

\newcommand{\DTubeProb}{0.833}
\newcommand{\DTubeTime}{89.8}

\newcommand{\DMatchedProbGain}{6.9}
\newcommand{\DMatchedProbGainLow}{6.7}
\newcommand{\DMatchedProbGainHigh}{7.0}
\newcommand{\DMatchedTimeSaving}{11.5}
\newcommand{\DMatchedTimeSavingLow}{11.4}
\newcommand{\DMatchedTimeSavingHigh}{11.7}

\newcommand{\KSpacecraftN}{83}

\newcommand{\FMaxSites}{24}

\newcommand{\RDelayTimeSaving}{5.1}

\title{Optimal Search for Finding Satellites Post-Launch}

\author{
Carlo Schreiber\textsuperscript{1} \hspace{1.5em}
Duncan Eddy\textsuperscript{1} \hspace{1.5em}
Mahdi Al-Husseini\textsuperscript{1}\\
Derek Woods\textsuperscript{2} \hspace{1.5em}
Araz Feyzi\textsuperscript{2} \hspace{1.5em}
Mykel J. Kochenderfer\textsuperscript{1}\\[.3em]
\begin{tabular}{@{}c@{\hspace{2em}}c@{}}
\textsuperscript{1}Stanford University & \textsuperscript{2}Kayhan Space\\
496 Lomita Mall & 10901 W 120th Ave \#140\\
Stanford, CA 94305 & Broomfield, CO 80021\\
\texttt{\{carlops, deddy, mah9, mykel\}@stanford.edu} &
\texttt{\{derek.woods, araz\}@kayhan.space}
\end{tabular}
\thanks{\footnotesize This work has been submitted to the IEEE for possible publication. Copyright may be transferred without notice, after which this version may no longer be accessible.}
}

\begin{document}

\maketitle
\thispagestyle{plain}
\pagestyle{plain}

\begin{abstract}

This paper introduces Bayesian Continuation Model Predictive Control (BC-MPC), an adaptive method that replans near-term antenna pointings while continuing a feasible baseline sweep, for finding and establishing contact with spacecraft immediately after launch when their true positions are most uncertain. After separation, launch and deployment errors can leave the spacecraft sufficiently far from its predicted position that operators must actively search for it during ground contacts to establish communications. Each missed detection provides information about which orbital hypotheses remain plausible, but searching one part of the uncertainty region can consume the time needed to reach another. BC-MPC updates a particle-based orbital belief after missed detections and replans antenna pointing guidance while retaining feasibility. Its continuation helps preserve future search coverage that greedy and short-horizon methods can sacrifice for immediate acquisition gains. We compare BC-MPC against the operationally prevalent along-track sweep and additional search algorithms under common search-time budgets, initial uncertainty, and antenna constraints. Across \DDeployments{} simulated deployments with matched uncertainty, BC-MPC improves acquisition probability over the along-track sweep and reduces restricted mean acquisition time. We additionally evaluate the method using pre-launch, post-launch, and initial acquisition ephemerides for \KSpacecraftN{} spacecraft from the Transporter-16 rideshare launch. Using either pre-deployment or post-deployment two-line element sets (TLEs) for planning, BC-MPC improves modeled acquisition probability over the along-track sweep by approximately 12 percentage points under the commonly assumed uncertainty model.

Website: https://satellite-acquisition.github.io/
\end{abstract}

\tableofcontents

\begin{figure*}[!t]
 \centering
 \includegraphics[width=0.95\textwidth]{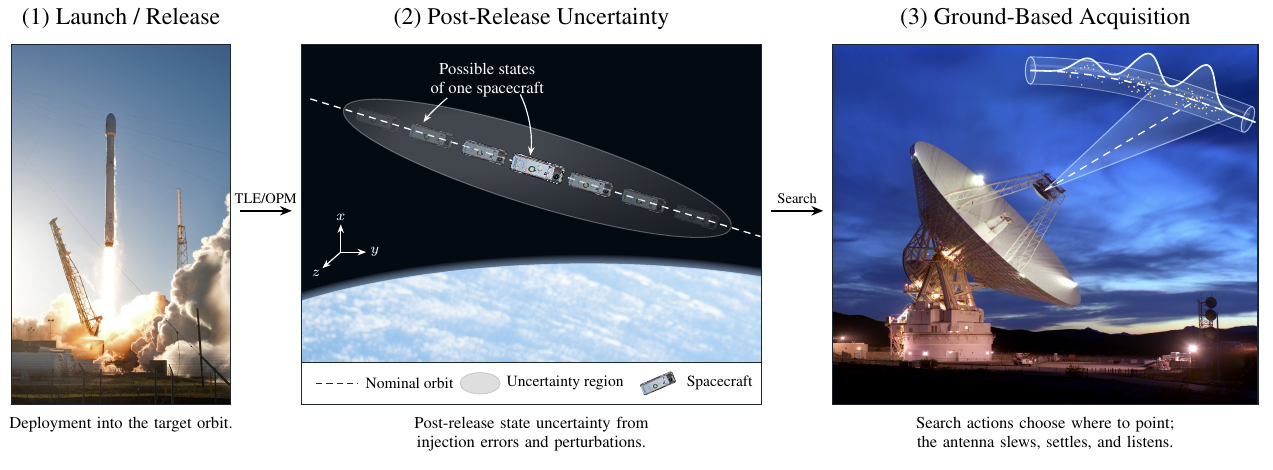}
 \caption{Satellite launch and initial acquisition problem:
 An orbit parameter message provides an
 initial orbit state estimate. Operators must then command antenna pointing offsets, followed by
 slewing, settling, and listening to search for the spacecraft. 
 Photo credits: NASA/Tony Gray (1) and NASA (2, 3).}
 \label{fig:acquisition-overview}
\end{figure*}

\section{Introduction}
\label{sec:introduction}

Launch and early-orbit operations are among the most challenging phases of a spacecraft mission. After launch, the spacecraft may have experienced a fault and operators must establish contact using ground-based antennas to assess spacecraft health and transition toward routine communications despite not knowing its exact position ~\cite{fernandezmartin1999skybridge,sieg2007nonnominal}. To make contact with spacecraft, operators rely on pre-launch orbit parameter messages (OPMs) from the launch vehicle provider, which contain the expected deployment state and inform the antenna pointing search strategy. The challenge of this search is exacerbated by uncertainty introduced by launch-vehicle injection errors, ejection-induced dispersion, and orbital perturbations, in particular drag. The resulting uncertainty can exceed the antenna beamwidth, requiring operators to search around the predicted spacecraft location rather than track a single nominal trajectory.

This search must be performed under tight operational constraints. Ground contacts are limited, and an antenna can search only one location at a time. The challenge is particularly acute for rideshare launches, when many spacecraft may require acquisition over overlapping passes. Search decisions must therefore use the available ground contact pass time efficiently while respecting antenna slew, settling, and communication timing constraints.

Operational acquisition procedures have traditionally searched along or around the predicted spacecraft trajectory using predetermined scan patterns \cite{distasio1999optimal,fernandezmartin1999skybridge}. Such procedures are simple and predictable, but they do not generally use missed detections to change where the antenna should search next.

Adapting the search involves two important considerations. First, each unsuccessful observation provides information. If the antenna points at a location where a particular orbit hypothesis should have produced a detection, a miss reduces the plausibility of that hypothesis. Second, antenna motion limits which locations remain reachable later in the contact. Moving toward one part of the uncertainty region can consume the time needed to reach another. An adaptive search strategy must therefore use information from missed detections without sacrificing coverage of the remaining uncertainty.

These characteristics connect spacecraft acquisition to classical search theory, which studies how search effort should be allocated over an uncertain target distribution \cite{koopman1956,stone2016moving}. Bayesian search methods provide a natural way to update uncertain distributions after unsuccessful observations \cite{koch2007negative}. Related work in space-object recovery has considered particle-based search \cite{hobson2013particle}, minimum-time sensor tasking \cite{fedeler2022minimumtime}, and information-driven control \cite{wolf2025infodriven}. Uncertainty-shaped optical scans and adaptive frequency sweeps likewise exploit structure in spatial and Doppler uncertainty \cite{yun2025elliptical,ortega2021adaptive}. Our focus is on combining this information-driven adaptation with the problem-specific search constraints imposed by antenna motion and finite contact duration.

We introduce Bayesian Continuation Model Predictive Control (BC-MPC) for spacecraft acquisition under orbital uncertainty. Model predictive control (MPC) repeatedly optimizes actions over a finite horizon, executes the first, and replans as new measurements arrive~\cite{mayne2000constrained}. BC-MPC maintains a particle-based belief over spacecraft trajectories that is updated after missed detections. Starting from an along-track uncertainty sweep, it replans the next antenna pointings while retaining the rest of the sweep as a feasible continuation, so each pointing is judged by acquisition probability over the full remaining contact rather than the horizon alone. The overall procedure is illustrated in \Cref{fig:acquisition-overview}.

The paper makes three contributions:
\begin{itemize}
    \item We formulate spacecraft acquisition as a sequential decision problem that combines orbital uncertainty, missed-detection feedback, and antenna-motion constraints.
    \item We introduce BC-MPC, a search controller that adapts near-term pointings while accounting for acquisition probability over the full remaining contact.
    \item We evaluate the method across \DDeployments{} simulated deployments, study sensitivity to antenna motion, repeated contacts, and fleet acquisition, and perform an operational-data evaluation on Transporter-16 rideshare deployment orbit data.
\end{itemize}

The paper proceeds as follows. Section 2 formulates the spacecraft acquisition problem. Section 3 introduces BC-MPC and the baseline search strategies. Section 4 evaluates single-spacecraft and fleet acquisition and presents the Transporter-16 replay. Section 5 concludes with directions for future work.

\section{Problem Formulation}
\label{sec:problem-formulation}

We consider the problem of selecting a sequence of antenna search actions \(a_{1:L}=(a_1,\ldots,a_L)\) that maximizes the probability of acquiring a given spacecraft within a ground-station contact window, given a belief \(\mathcal{B}\) about its state \(x\). Each action commands an antenna pointing, followed by settling and listening for a detection. This formulation contains four components: the contact window, the orbital belief, the search actions and observations, and the antenna motion constraints.

\subsection{Contact windows}

For a given ground station, a contact window \(c(x)=[t_s(x),t_f(x)]\) is the interval during which the antenna can attempt to communicate with a spacecraft, beginning at \(t_s\) and ending at \(t_f\). These times depend on the spacecraft state \(x\) and the station geometry, and are determined by when the spacecraft rises above and falls below the station's elevation mask. Because the true state is not known precisely after launch, planning uses a predicted window \(\hat c=c(\hat x)\), where \(\hat x\) is the nominal state supplied by the available orbit product, around which the initial belief \(\mathcal{B}\) is constructed. The actual visibility interval may therefore differ from the predicted window. From the initial belief, we predict a set of contact windows and plan pointing actions within each selected window. If acquisition fails, we wait until the next available contact and resume with the updated belief.

\subsection{Orbital uncertainty}

Launch providers supply orbital estimates through pre-deployment and post-deployment OPMs. A pre-deployment OPM is based on the planned launch trajectory, while a post-deployment OPM incorporates as-flown telemetry to account for deviations during ascent. These products provide a nominal orbit, but the true spacecraft state remains uncertain because of launch and deployment errors, limited ground-based tracking data, and orbit prediction error. An accompanying uncertainty model is therefore needed to describe where the spacecraft may actually be \cite{ccsds_odm}. We represent this uncertainty using a weighted set of \(N\) orbit hypotheses, or particles, \(\{(x_i,w_i)\}_{i=1}^{N}\), with nonnegative weights that sum to one. Each particle represents one possible spacecraft trajectory and is propagated through the full contact window using the Simplified General Perturbations 4 (SGP4) orbit propagator~\cite{vallado2006sgp4}.

For a ground station \(g\), a spacecraft state \(x\) predicts an antenna pointing direction at time \(t\) through the mapping
\begin{equation}
    q=f_g(x,t), \qquad q\in\mathbb{S}^2.
    \label{eq:state-to-pointing}
\end{equation}
This mapping propagates the state to \(t\), transforms the spacecraft position into Earth-fixed coordinates, subtracts the station position, and expresses the resulting line of sight in the station's local topocentric frame. Normalizing this vector gives the pointing direction \(q\), equivalently represented by azimuth and elevation. Each orbital hypothesis therefore has an associated antenna pointing direction \(q_i(t)=f_g(x_i,t)\).

\subsection{Search actions and observations}

At decision step \(k\), the operator executes a search action \(a_k\), consisting of commanding the antenna to a pointing direction \(q_k\in\mathbb{S}^2\), letting it settle, and then listening for the spacecraft. Under hypothesis \(i\), the pointing error at time \(t\) is the angular separation between the commanded direction \(q_k\) and the predicted spacecraft direction \(q_i(t)\). A communications model uses this error, together with range, elevation, and listening duration, to determine the acquisition probability \(p_i(a_k)\). Specifically, the one-second detection probability \(p_{1,i}(a_k)\) follows a Gaussian profile in angular pointing error, with a full width at half maximum of \qty{2}{\degree} and an on-axis value that decreases with increasing range and decreasing elevation. For a listening duration \(d\), we compute \(p_i(a_k)=1-[1-p_{1,i}(a_k)]^{d/(\qty{1}{\second})}\), holding geometry fixed at a representative time within the listening interval. In the experiments considered here, each action produces a binary outcome: a confirmation or a miss. A confirmation terminates the search, while a miss allows the search to continue. For a generic action \(a\), we denote the probability of a miss under hypothesis \(i\) by
\begin{equation}
     s_i(a) = 1 - p_i(a).
\end{equation}

After a miss, the particle weights are updated using Bayes' rule,
\begin{equation}
    w_i^{+}
    =
    \frac{w_i s_i(a)}
         {\sum_j w_j s_j(a)}.
    \label{eq:bayes-update}
\end{equation}

A missed detection therefore shifts probability away from trajectories for which the executed pointing was likely to result in acquisition, and toward trajectories that are more consistent with the miss. For a given spacecraft trajectory and pointing sequence, we treat successive detection attempts as independent, ignoring effects such as persistent propagation losses or beacon outages.

For a sequence of \(L\) search actions \(a_{1:L}\), the total probability of acquisition is
\begin{equation}
    J(a_{1:L})
    =
    \sum_i w_i
    \sum_{h=1}^{L}
    p_i(a_h)
    \prod_{\ell=1}^{h-1} s_i(a_\ell).
    \label{eq:acquisition-objective}
\end{equation}

\subsection{Antenna motion feasibility}

Search actions are constrained by the physical motion of the ground-station antenna. Starting from an initial pointing \(q_0\), successive actions require transfers from \(q_{k-1}\) to \(q_k\). We model each slew as a great-circle transfer that begins and ends at rest. If \(\theta(t)\) denotes angular progress along this transfer, the mount limits require
\begin{equation}
    |\dot\theta(t)|\leq\omega_{\max},
    \qquad
    |\ddot\theta(t)|\leq\alpha_{\max},
\end{equation}
where \(\omega_{\max}\) and \(\alpha_{\max}\) are the maximum slew rate and angular acceleration. Let \(T_{\mathrm{slew}}(q_{k-1},q_k)\) be the minimum transfer time under these limits. A pointing is reachable within a decision interval \(\Delta t\) only if
\begin{equation}
    T_{\mathrm{slew}}(q_{k-1},q_k)
    +t_{\mathrm{settle}}+t_{\mathrm{listen}}
    \leq\Delta t,
    \label{eq:motion-feasibility}
\end{equation}
where \(t_{\mathrm{settle}}\) and \(t_{\mathrm{listen}}\) are the settling and listening durations. Thus, selecting one pointing changes the starting point for the next slew and can restrict which search actions remain feasible later in the contact window.

\section{Bayesian Continuation MPC}
\label{sec:search-mpc}

Operators can use several search strategies to acquire a spacecraft when the predicted orbit is uncertain. As the spacecraft moves along its orbit, its predicted position changes continuously throughout the contact window. The search strategies considered here therefore define antenna pointings relative to this moving nominal trajectory. We first describe common operational strategies before introducing Bayesian Continuation Model Predictive Control.

\subsection{Operational search strategies}

The most direct approach is \emph{nominal tracking}~\cite{slobin2015positioning}, which commands the pointing direction predicted by the nominal spacecraft state \(\hat x\):
\begin{equation}
    q_{\mathrm{nom}}(t)=f_g(\hat x,t).
\end{equation}
When the orbit prediction is sufficiently accurate, nominal tracking can acquire the spacecraft quickly without requiring an explicit search. Nominal tracking is not robust to any errors or uncertainty in the initial position. Therefore a slightly more advanced search strategy is a grid sweep. A uniform grid sweep searches a fixed region around the predicted spacecraft position in both the along-track and cross-track directions, without exploiting the structure of the orbital uncertainty. While this provides broader coverage than nominal tracking, operators may spend substantial time searching locations where the spacecraft is unlikely to be.

\begin{figure}[H]
    \centering
    \includegraphics[width=0.9\columnwidth]{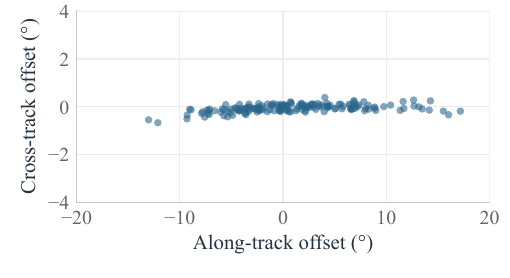}
    \caption{Angular uncertainty for a simulated
    contact. Dots show planning particles; axis scales differ.}
    \label{fig:belief-geometry}
\end{figure}

A better operational approach is an Along-Track Uncertainty Sweep (ATUS). Because post-deployment orbital uncertainty is typically elongated along the direction of motion~\cite{jonglez2022tubin}, ATUS concentrates the search along this dominant uncertainty direction (\Cref{fig:belief-geometry}). We define the ATUS policy used here as a fixed, repeating sweep of the propagated prior, subject to antenna motion limits. ATUS uses \(2m+1\) candidate pointings, indexed from \(-m\) to \(m\), evenly spaced along the predicted along-track uncertainty, with index \(0\) at its center. The preferred pointing follows the repeating sequence
\begin{equation}
    0,1,\ldots,m,\;m-1,\ldots,-m,\;-m+1,\ldots,-1.
    \label{eq:atus-order}
\end{equation}
The candidate directions move with the predicted trajectory. At each step, the antenna selects the reachable candidate closest to the preferred pointing, or makes a partial slew toward it if none is reachable. The sweep advances one index per step; missed detections do not alter its order.

Lastly, a Bayesian greedy strategy instead uses the updated orbital belief after a given position is considered a miss and points the antenna toward the feasible location with the highest immediate probability of acquisition. This makes the search adaptive, but considers only the next pointing and does not account for how that decision affects later search opportunities.

\subsection{Bayesian Continuation MPC}

Looking beyond a single pointing leads naturally to model predictive control (MPC). Rather than selecting only the next action, MPC optimizes a sequence of \(H\) upcoming antenna pointings using the current orbital belief. Only the first pointing is executed before the optimization is repeated. 

A short planning horizon may cover only part of the remaining contact window, leaving later search opportunities outside the current optimization. BC-MPC addresses this limitation by retaining an ATUS continuation beyond the MPC horizon. Starting from an ATUS schedule, BC-MPC reoptimizes the next \(H\) antenna pointings while keeping the remainder of the sweep unchanged. The revised pointings are therefore evaluated together with the retained continuation over the full remaining contact window. This construction is closely related to rollout methods in approximate dynamic programming~\cite{bertsekas1997rollout} and MPC with terminal costs and policies~\cite{mayne2000constrained}, which account for downstream performance beyond the optimized horizon. In BC-MPC, the retained, motion-feasible search schedule provides this continuation, evaluated under the current Bayesian belief.

Suppose \(K\) search stages remain, with \(1\leq H\leq K\), and let \(\bar a_{1:K}\) denote the current feasible action schedule, initialized by ATUS as defined above. For each candidate prefix \(a_{1:H}\), form a complete schedule by retaining the fixed continuation:
\begin{equation}
\tilde a_j(a_{1:H})=
\begin{cases}
a_j, & 1\leq j\leq H,\\
\bar a_j, & H<j\leq K.
\end{cases}
\label{eq:bcmpc-schedule}
\end{equation}

BC-MPC optimizes only the prefix, but evaluates the probability of missing the spacecraft over all \(K\) remaining stages:
\begin{equation}
a_{1:H}^{\star}
=
\arg\min_{a_{1:H}}
\sum_{i=1}^{N}
w_i
\prod_{j=1}^{K}
s_{i,j}\!\left(\tilde a_j(a_{1:H})\right),
\label{eq:bcmpc-objective}
\end{equation}
where \(s_{i,j}(a)\) is the miss probability for hypothesis \(i\) under action \(a\) at remaining stage \(j\). The complete schedule must satisfy the antenna motion constraints, including the transition from the optimized prefix to the retained continuation.

The optimized schedule is \(\tilde a_{1:K}(a_{1:H}^{\star})\): its first \(H\) actions are \(a_{1:H}^{\star}\), and its remaining actions are \(\bar a_{H+1:K}\). Only \(a_1^{\star}\) is executed. If the spacecraft is acquired, the search terminates. After a miss, the particle weights are updated using \Cref{eq:bayes-update}, the optimized schedule is shifted by one stage and retained as the next continuation, and the optimization is repeated using the updated belief.

\section{Evaluation}
\label{sec:evaluation}

To evaluate BC-MPC, we focus on two main areas. First, we study the controller's effect on single-spacecraft acquisition, including sensitivity to antenna performance, orbital uncertainty, and receiver behavior. Second, we evaluate how contact-level search improvements translate to fleet acquisition across repeated contact opportunities. Finally, we evaluate BC-MPC through an operational-data replay using operational orbit data from \KSpacecraftN{} spacecraft from the Transporter-16 rideshare launch provided by Kayhan Space.

Performance is evaluated using acquisition probability and restricted mean acquisition time (RMST). Let \(T\) be the elapsed time from the start of a search until acquisition is confirmed, and let \(\tau\) be the allocated search-window duration. Acquisition probability is the probability that confirmation occurs within this window, while RMST is
\begin{equation}
    \operatorname{RMST}(\tau)
    =\mathbb{E}\!\left[\min\{T,\tau\}\right].
    \label{eq:rmst}
\end{equation}
Thus, acquisition time is capped before averaging: successful searches contribute their time to confirmation, and searches without confirmation within the window contribute the full duration \(\tau\).

\begin{figure*}[!t]
    \centering
    \includegraphics[width=0.9\textwidth]{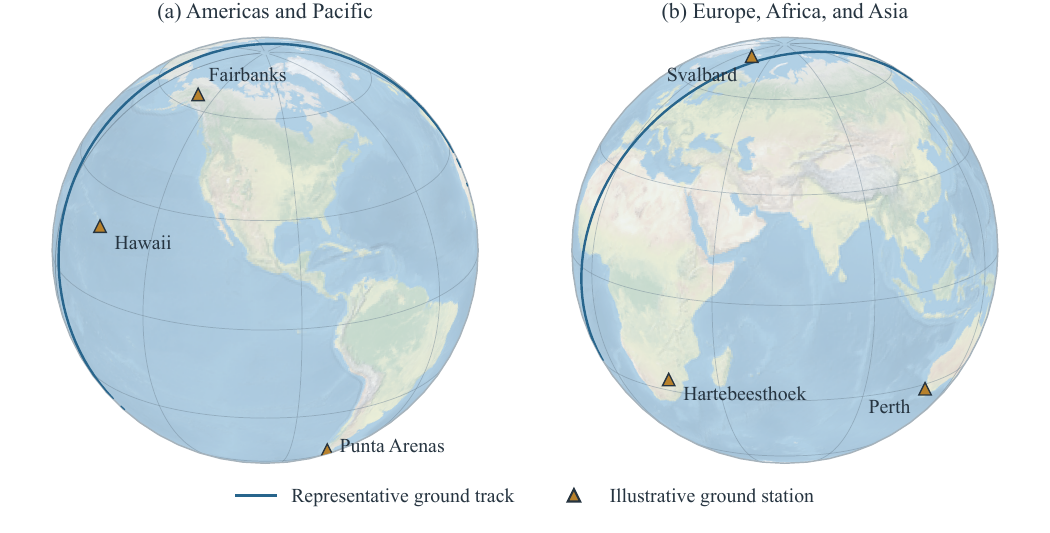}
    \caption{Illustrative six-station ground network and an example spacecraft orbit used in the evaluation.}
    \label{fig:deployment-globe}
\end{figure*}

The primary evaluation uses six illustrative ground stations (\Cref{fig:deployment-globe}): Svalbard, Fairbanks, Hawaii, Hartebeesthoek, and Punta Arenas from KSAT's network~\cite{ksat_network_map}, plus Perth, Australia. We use approximate coordinates and one independently steerable antenna per station. Orbits are propagated with Brahe astrodynamics library~\cite{eddy2026brahe} and SGP4~\cite{vallado2006sgp4}. For the primary evaluation and Transporter-16 replay, equally weighted particles are initialized by independent Gaussian perturbations about the nominal orbit, with assumed standard deviations of \qty{20}{\second} in orbital phase, \qty{0.025}{\degree} in inclination, \qty{0.035}{\degree} in right ascension of the ascending node, and \num{2.5e-4} revolutions per day in mean motion. \Cref{tab:evaluation-design} summarizes the shared search configuration.

\begin{table}[H]
    \centering
    \caption{Parameters used for simulation.}
    \label{tab:evaluation-design}
    \small
    \begin{tabular}{@{}p{0.53\columnwidth}@{\hspace{0.035\columnwidth}}p{0.40\columnwidth}@{}}
        \toprule[1pt]
        Parameter & Setting \\
        \midrule[0.8pt]
        Search window & \qty{192}{\second} \\
        Move / settle / listen per stage & \qty{4.0}{\second} / \qty{0.5}{\second} / \qty{1.5}{\second} \\
        Planning / evaluation particles & 160 / 640 \\
        Candidate pointings / extent & 17 / $\pm2.8$ along-track SD \\
        MPC horizon / beam width & 4 stages / 40 \\
        Antenna rate / acceleration & \qty{1.10}{\degree\per\second} / \qty{1.00}{\degree\per\second\squared} \\
        Beam FWHM / elevation mask & \qty{2}{\degree} / \qty{10}{\degree} \\
        \bottomrule[0.8pt]
    \end{tabular}
\end{table}

\subsection{Single-spacecraft search}
\label{sec:synthetic-evaluation}

We evaluate first-contact acquisition across \DAttempts{} attempts from \DDeployments{} realizations of a spacecraft deployment near an orbit with an altitude of \qty{550}{\kilo\metre}
and an inclination of \qty{97.6}{\degree}, with spacecraft elements varied across deployments to produce different acquisition geometries. BC-MPC is initialized from an ATUS schedule with a planning horizon of \(H=4\) dwells and compared against a uniform grid sweep, ATUS, a Bayesian greedy controller, and an MPC without its continuation term. \Cref{tab:deployment-results} summarizes the results.

\begin{table}[H]
    \centering
    \caption{First-contact acquisition probability and mean acquisition time.}
    \label{tab:deployment-results}
    \small
    \begingroup
    \setlength{\heavyrulewidth}{1pt}
    \setlength{\lightrulewidth}{0.8pt}
    \begin{tabular*}{0.96\columnwidth}{@{\extracolsep{\fill}}lccc@{}}
\toprule Controller & $P_{60}$ & $P_{192}$ & RMST (s) \\ \midrule
Uniform grid & 0.248 & 0.329 & 141.9 \\
ATUS & 0.376 & 0.833 & 89.8 \\
Bayesian greedy & 0.565 & 0.688 & 86.5 \\
MPC, no continuation & \textbf{0.614} & 0.790 & 74.5 \\
BC-MPC & 0.423 & \textbf{0.902} & 78.3 \\
\bottomrule \end{tabular*}

    \endgroup
\end{table}

Over the full contact window of \qty{192}{\second}, BC-MPC achieves the highest acquisition probability of \DMpcProb{}, compared with \DTubeProb{} for ATUS, corresponding to an improvement of \DMatchedProbGain{} percentage points (95\% confidence interval \DMatchedProbGainLow{}--\DMatchedProbGainHigh{}). BC-MPC also reduces RMST from \qty{\DTubeTime}{\second} for ATUS to \qty{\DMpcTime}{\second}, a reduction of \qty{\DMatchedTimeSaving}{\second} (95\% interval \DMatchedTimeSavingLow{}--\DMatchedTimeSavingHigh{}). 

\begin{figure}[H]
    \centering
    \includegraphics[width=0.95\columnwidth]{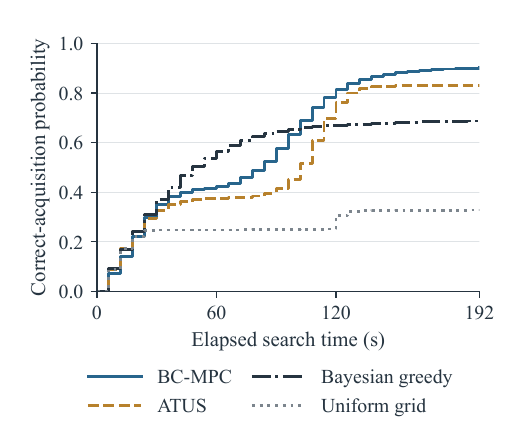}
    \caption{First-contact acquisition probability over elapsed search time.}
    \label{fig:deployment-results}
\end{figure}

An interesting finding, illustrated in \Cref{fig:deployment-results}, is that strategies with higher acquisition probability early in the search can have lower acquisition probability over the full contact window. By \qty{60}{\second} of elapsed search time, Bayesian greedy and MPC without continuation both achieve higher cumulative acquisition probabilities than BC-MPC. However, by the end of the \qty{192}{\second} search window, BC-MPC achieves the highest acquisition probability among the displayed controllers. This tradeoff highlights the value of retaining effective search opportunities for later in the contact when early pointings do not yield a detection.

\begin{figure}[H]
    \centering
    \includegraphics[width=0.9\columnwidth]{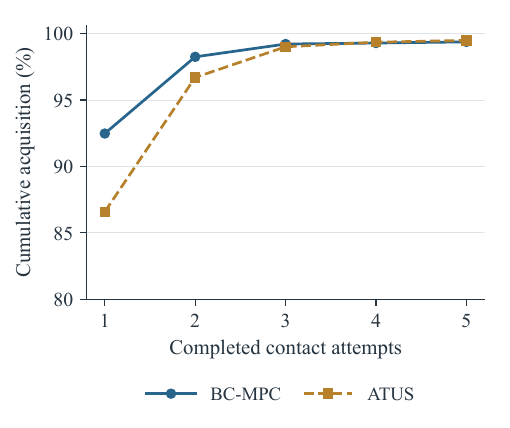}
    \caption{Cumulative acquisition across repeated contact attempts for
    BC-MPC and ATUS.}
    \label{fig:contact-history}
\end{figure}

If the search is unsuccessful over an entire contact window, the unsuccessful observations provide useful information about the spacecraft trajectory. Subsequent contacts therefore begin with a more informed belief, allowing adaptive search decisions to better exploit the remaining uncertainty. We evaluate how this accumulated information affects acquisition over repeated contact opportunities across 16 spacecraft scenarios by comparing cumulative acquisition for BC-MPC and ATUS in \Cref{fig:contact-history}. BC-MPC acquires more spacecraft during the earliest contacts, while the difference narrows as additional opportunities become available.

\begin{figure}[H]
    \centering
    \includegraphics[width=0.9\columnwidth]{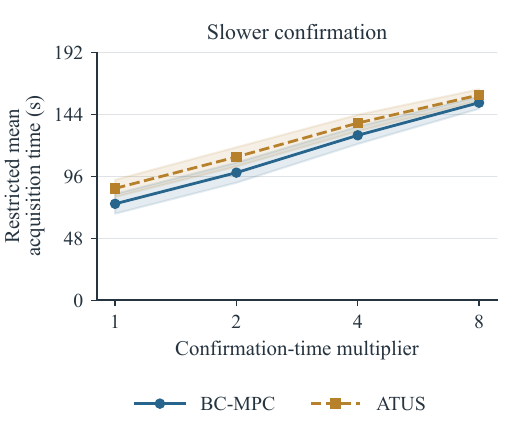}
    \caption{RMST with slower receiver confirmation.
    Bands show pointwise 95\% confidence intervals.}
    \label{fig:receiver-confirmation}
\end{figure}

We next evaluate how slower or delayed satellite responses affect acquisition performance, as might occur when spacecraft tumbling affects antenna gain. First, we increase the time required for acquisition confirmation while keeping the controller search plans unchanged. As shown in \Cref{fig:receiver-confirmation}, RMST increases for both BC-MPC and ATUS as confirmation becomes slower, while BC-MPC consistently maintains a lower RMST than ATUS across the tested confirmation-time multipliers.

\begin{figure}[H]
    \centering
    \includegraphics[width=\columnwidth]{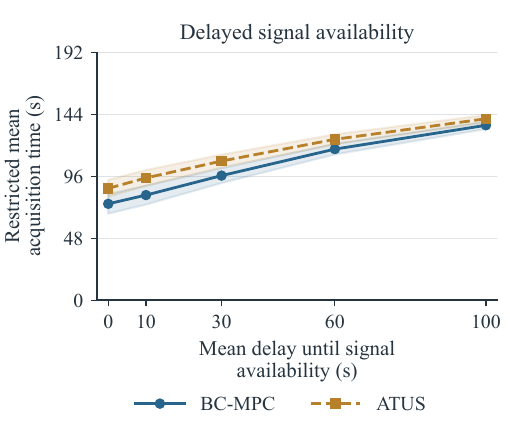}
    \caption{RMST with delayed signal availability.
    Bands show pointwise 95\% confidence intervals.}
    \label{fig:receiver-delay}
\end{figure}

The timing advantage also persists when signal availability is delayed, and not available until some time into the contact window (\Cref{fig:receiver-delay}), as might occur with a miscalibrated horizon mask or a software configuration issue. As the mean delay increases, RMST increases for both controllers, while the advantage of BC-MPC over ATUS gradually narrows as signal availability increasingly limits acquisition. For a positive mean delay, every target starts unavailable and becomes available after an exponentially distributed wait. Even at a \qty{100}{\second} mean signal-availability delay, BC-MPC retains an average RMST saving of \qty{\RDelayTimeSaving}{\second}.

Finally, we examine how antenna motion and MPC planning horizon affect acquisition performance. We first vary antenna agility in a supporting study. \Cref{tab:mechanism-controllers} compares ATUS, Bayesian greedy, and BC-MPC under constrained and agile antenna dynamics across eight scenarios.

\begin{table}[H]
    \centering
    \caption{Acquisition by \qty{192}{\second} (\%).
    Entries are means with 95\% normal intervals.}
    \label{tab:mechanism-controllers}
    \small
    \begingroup
    \renewcommand{\arraystretch}{1.22}
    \setlength{\heavyrulewidth}{1pt}
    \setlength{\lightrulewidth}{0.8pt}
    \begin{tabular*}{0.99\columnwidth}{@{\extracolsep{\fill}}lcc@{}}
\toprule
Controller & Slower antenna (1.1$^\circ$/s) & Agile antenna (8$^\circ$/s) \\
\midrule
ATUS & 76.1 [73.2, 79.0] & 82.7 [80.3, 85.2] \\
Bayesian greedy & 66.2 [63.5, 68.9] & 97.4 [97.0, 97.8] \\
BC-MPC & \textbf{90.9} [88.6, 93.1] & \textbf{97.9} [97.5, 98.2] \\
\bottomrule
\end{tabular*}
    \endgroup
\end{table}

BC-MPC provides a substantially larger advantage over Bayesian greedy when antenna motion is constrained to \qty{1.1}{\degree\per\second}. With an agile antenna (\qty{8}{\degree\per\second}) maximum speed), however, Bayesian greedy approaches the performance of BC-MPC. In this regime, rapid antenna motion keeps most future pointings reachable, so an immediate acquisition decision is less likely to restrict later search opportunities.

\begin{figure}[H]
    \centering
    \includegraphics[width=0.9\columnwidth]{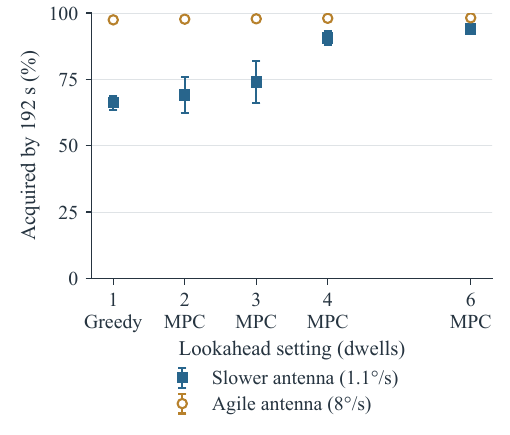}
    \caption{Acquisition by \qty{192}{\second} versus MPC horizon, without continuation.
    Whiskers show exploratory 95\% normal intervals.}
    \label{fig:mechanism-horizon}
\end{figure}

We further investigate the effect of greedy versus model predictive control by varying the MPC lookahead horizon. \Cref{fig:mechanism-horizon} shows that, with constrained antenna dynamics, increasing the number of lookahead dwells substantially improves terminal acquisition relative to greedy search. With an agile antenna, performance changes little as the horizon increases, confirming the behavior observed in ~\Cref{tab:mechanism-controllers}.

\subsection{Fleet acquisition after launch}
\label{sec:fleet-evaluation}

We next investigate whether contact-level improvements translate to improved fleet-scale acquisition. Such deployments release many spacecraft within a short period, creating competing acquisition demands across a limited ground-station network. We consider a fleet of 100 spacecraft all deployed from the same launch and measure the fraction acquired within the first 24 hours after deployment. We vary the number of available ground-station sites to evaluate how BC-MPC and ATUS perform as the number of acquisition opportunities increases.

\begin{figure}[H]
 \centering
 \includegraphics[width=0.9\columnwidth]{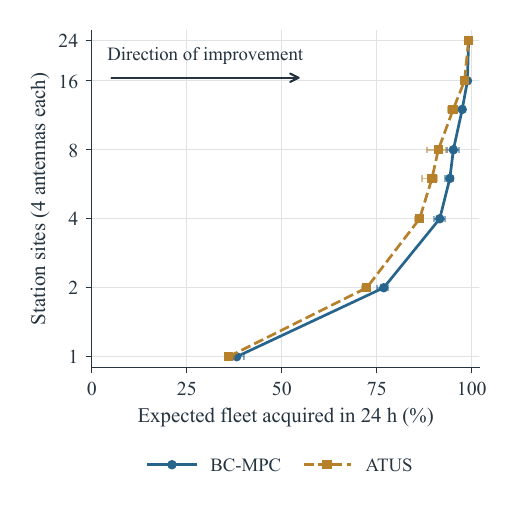}
 \caption{Fleet acquisition as the ground-station network expands to \FMaxSites{} sites. Whiskers show deployment ranges.}
 \label{fig:fleet-network}
\end{figure}

BC-MPC improves fleet-scale acquisition across the tested ground-station networks (\Cref{fig:fleet-network}). With six sites, the expected fraction of spacecraft acquired within 24 hours increases from 89.5\% with ATUS to 94.2\% with BC-MPC. The advantage is largest when ground-station opportunities are limited, where a missed acquisition is more difficult to recover from. As the network expands, additional contact opportunities reduce the impact of individual misses, and the performance gap narrows, with both controllers approaching 99\% acquisition for the largest tested networks.
\subsection{Kayhan Space LEOP replay}
\label{sec:operational-leop}

Finally, we evaluate BC-MPC using operational orbit data supplied by Kayhan Space for \KSpacecraftN{} objects from the Transporter-16 rideshare launch on March 30, 2026. The dataset contains pre-deployment ephemeris distributed to operators three days before launch, post-deployment ephemeris, and the first cataloged TLE correlated to each deployment. The pre- and post-deployment ephemeris represent the same orbit products used operationally to cue early ground contacts, while the first-catalog TLE is used only retrospectively as an approximate trajectory reference. We evaluate two planning cases, using either the pre- or post-deployment ephemeris as the controller input and retain the same six ground stations and search configuration used in the previous evaluation. Results are summarized in
Table~\ref{tab:kayhan-results}.

\begin{table}[H]
    \centering
    \caption{Modeled single-pass acquisition and RMST using deployment ephemeris
    from the Transporter-16 replay.}
    \label{tab:kayhan-results}
    \small
    \begingroup
    \setlength{\heavyrulewidth}{1pt}
    \setlength{\lightrulewidth}{0.8pt}
    \renewcommand{\arraystretch}{1.15}
    \begin{tabular*}{0.99\columnwidth}{@{\extracolsep{\fill}}lcc@{}}
\toprule
Controller & Acquisition (\%) & RMST (s) \\
\midrule
\multicolumn{3}{l}{\textit{Planning input: Pre-deployment Ephemeris}} \\
Nominal tracking & 4.96 & 184.24 \\
ATUS             & 69.51 & 144.05 \\
BC-MPC           & \textbf{81.55} & \textbf{139.10} \\
\addlinespace[2pt]
\multicolumn{3}{l}{\textit{Planning input: Post-deployment Ephemeris}} \\
Nominal tracking & 6.03 & 182.67 \\
ATUS             & 68.78 & 144.42 \\
BC-MPC           & \textbf{81.24} & \textbf{139.16} \\
\bottomrule
\end{tabular*}
    \endgroup
\end{table}

The pre- and post-deployment planning cases yield very similar results, with BC-MPC improving modeled acquisition over ATUS by approximately 12 percentage points in both cases while also reducing RMST. This mirrors the trend observed in the synthetic evaluation and shows that the improvement persists when using operational orbit products.

\section{Conclusion}
\label{sec:conclusion}

This paper studies spacecraft acquisition under orbital uncertainty, where missed detections provide information about the spacecraft location while antenna motion constrains which search actions remain reachable. We introduced Bayesian Continuation Model Predictive Control (BC-MPC), which adapts antenna pointings after missed detections. Across 96 simulated deployments, BC-MPC increased acquisition probability from 83.3\% to 90.2\% and reduced RMST from \qty{89.8}{\second} to \qty{78.3}{\second} relative to ATUS. Repeated-contact and fleet experiments show that these contact-level improvements translate into earlier fleet acquisition, with the advantage narrowing as additional ground-station opportunities become available. In a replay using real pre- and post-deployment ephemeris provided by Kayhan Space for 83 spacecraft from Transporter-16, BC-MPC improved modeled acquisition over ATUS by 12.0 and 12.4 percentage points, respectively, under the commonly assumed uncertainty model, while also reducing RMST. Together, these results show how Bayesian belief updates and future antenna reachability can be combined to improve early-orbit spacecraft acquisition while retaining the structure of an operational sweep. Future work could validate the approach using GNSS measurements from the vehicle from the time of deployment for greater accuracy, and extend the controller to variable listening times, richer receiver feedback including partial signal strength, and reassignment of reservations after early acquisition.

\bibliographystyle{IEEEtran}
\begingroup\interlinepenalty=10000
\bibliography{references}
\endgroup

\thebiography

\begingroup
\setkeys{Gin}{keepaspectratio}

\begin{biographywithpic}{Carlo Schreiber}{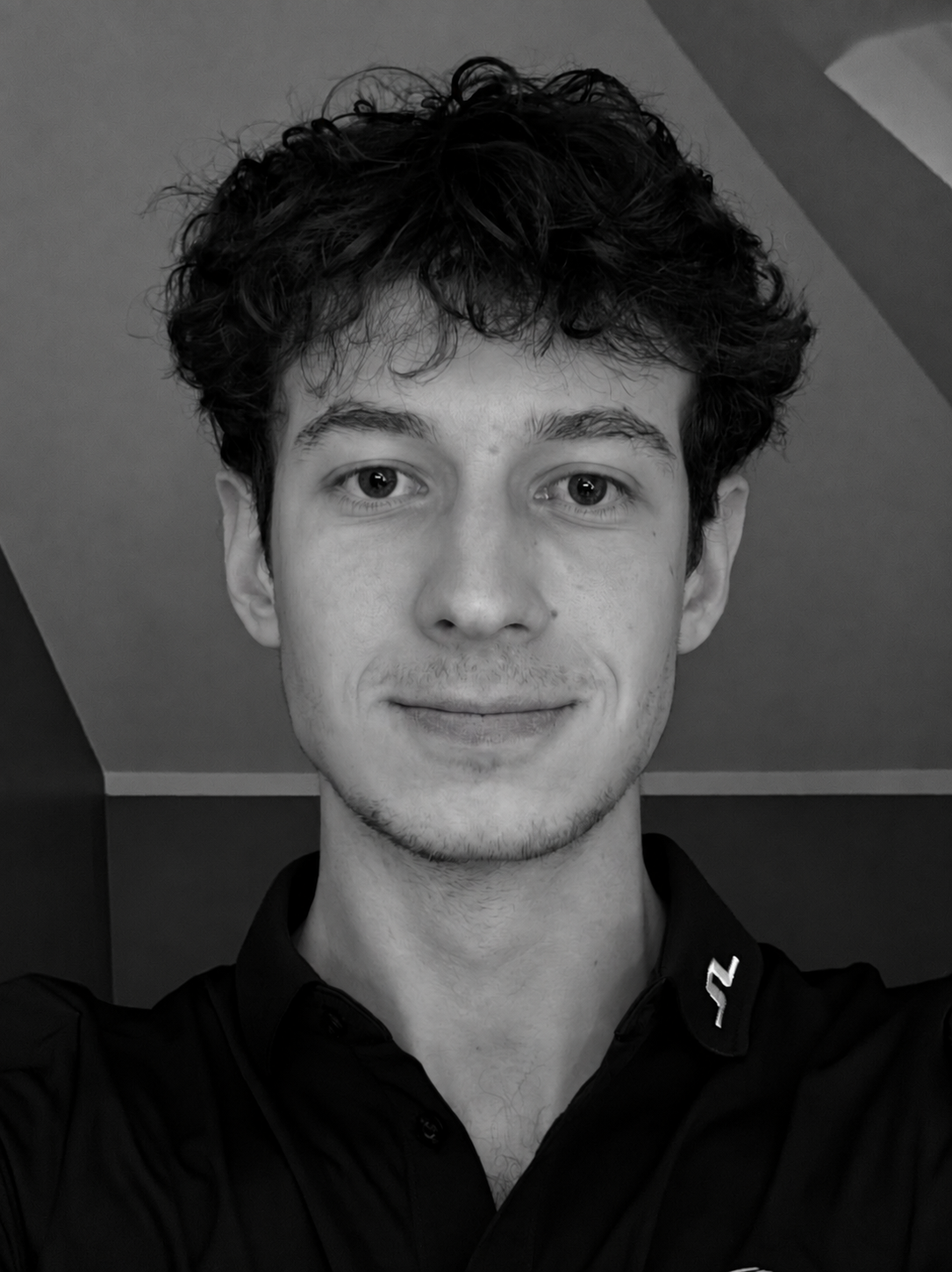} is an M.S. student in Electrical
Engineering at Stanford University and a graduate student researcher in the Stanford
Intelligent Systems Laboratory (SISL). He received his B.Sc. in Mechanical
Engineering from ETH Zurich, specializing in robotics and mechatronics.
His research focuses on methods for decision making under uncertainty, planning,
and control in autonomous systems, with a particular interest in aerospace
and robotics.
\end{biographywithpic}

\begin{biographywithpic}{Duncan Eddy}{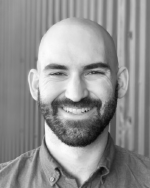}
is a staff research fellow in the Stanford Intelligent Systems Laboratory
(SISL) and the Executive Director of the Center for AI Safety at Stanford
University. His research focuses on decision-making in safety-critical,
climate, and space systems. He received a B.S. in Mechanical Engineering from
Rice University in 2013 and a Ph.D. in Aerospace Engineering from Stanford
University in 2021. Prior to returning to Stanford, he was the Director of
Space Operations at Capella Space Corporation, where he built a fully
automated constellation tasking and delivery system. He later founded and
led the Constellation Management and Space Safety Organization at Amazon's
Project Kuiper.
\end{biographywithpic}

\begin{biographywithpic}{Mahdi Al-Husseini}{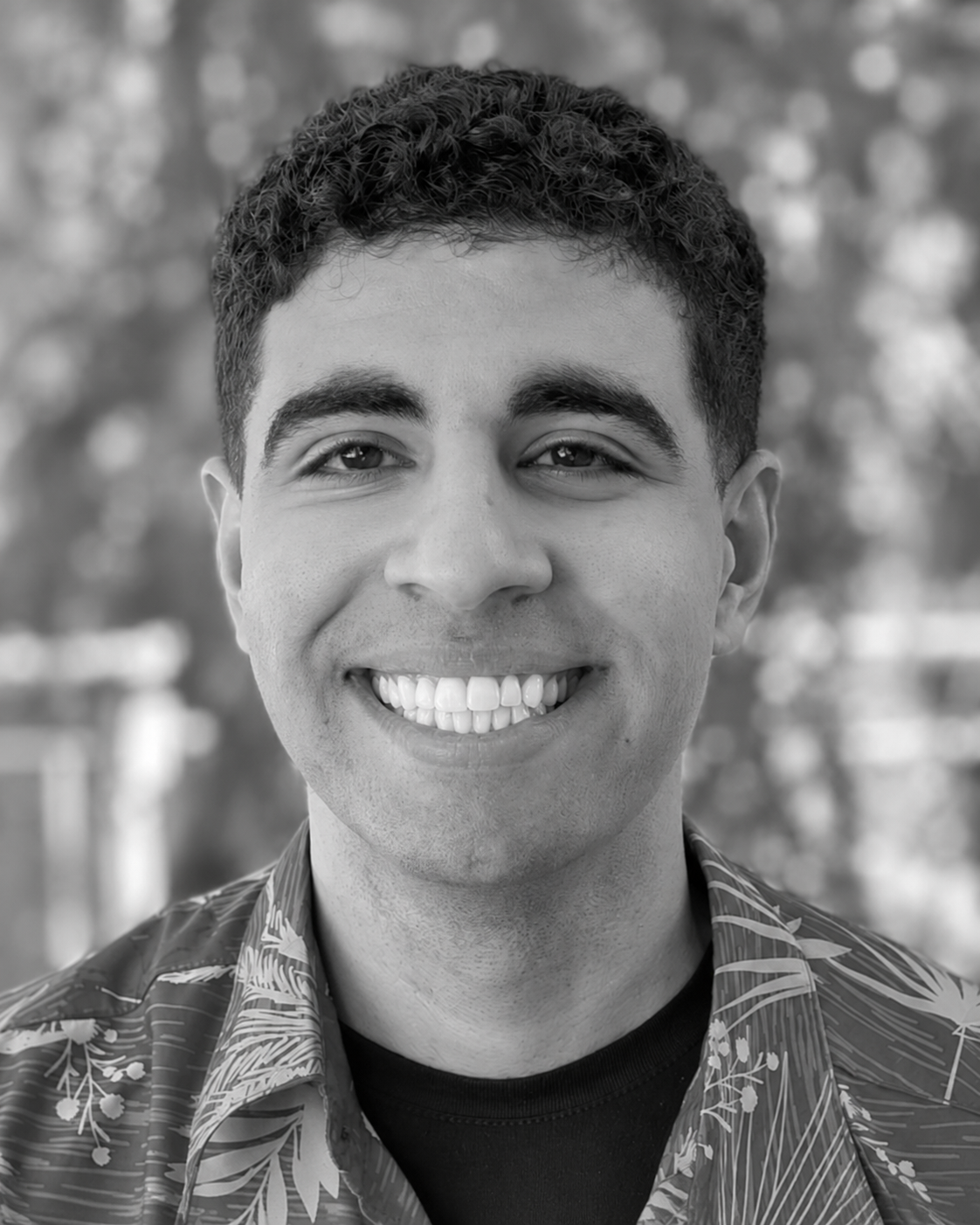}
is a Ph.D. candidate in the Stanford Intelligent Systems Laboratory (SISL).
His research develops models and algorithms for semi-decentralized multi-agent
systems operating under communication constraints. He received B.S. degrees
in Biomedical Engineering and Public Policy and an M.S. in Computer Science
from the Georgia Institute of Technology, and an M.S. in Aeronautics and
Astronautics from Stanford University. He is a U.S. Army captain and an
experimental test pilot in training.
\end{biographywithpic}

\begin{biographywithpic}{Derek Woods}{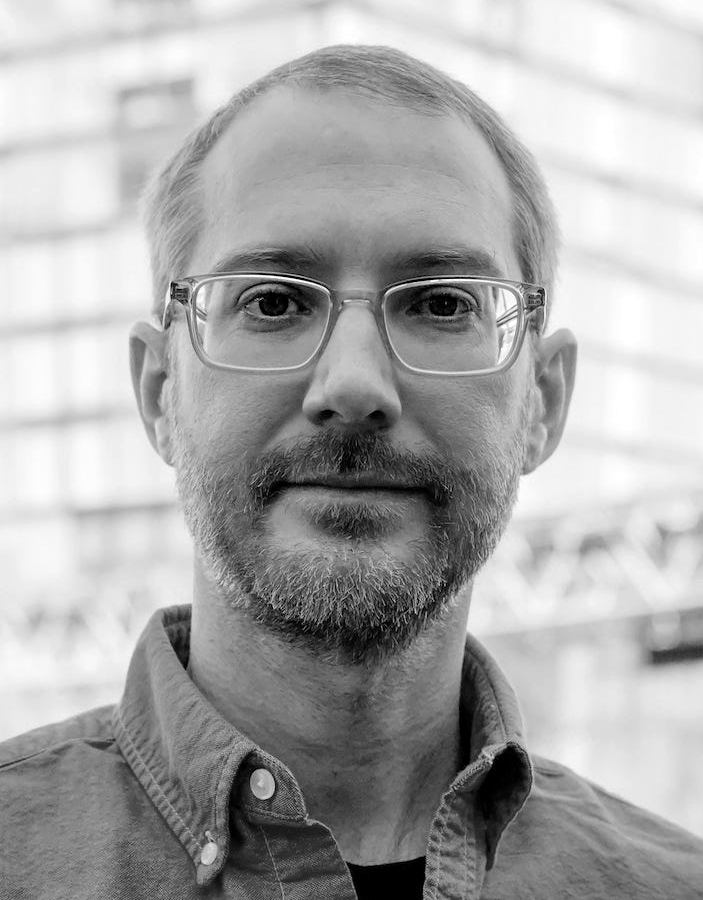}, Customer Success Manager, manages the integration of Kayhan Space’s spaceflight safety solutions for satellite operators, focusing on operational orbit determination, conjunction assessment, and Launch and Early Orbit Phase (LEOP) support. Previously, he served as a Senior Astrodynamics Engineer at Kayhan Space, building robust astrodynamics libraries centered on high-accuracy force modeling and code reliability. He holds an M.Sc. in Aerospace Engineering from the University of Florida, where he specialized in orbital mechanics and control theory.
\end{biographywithpic}

\begin{biographywithpic}{Araz Feyzi}{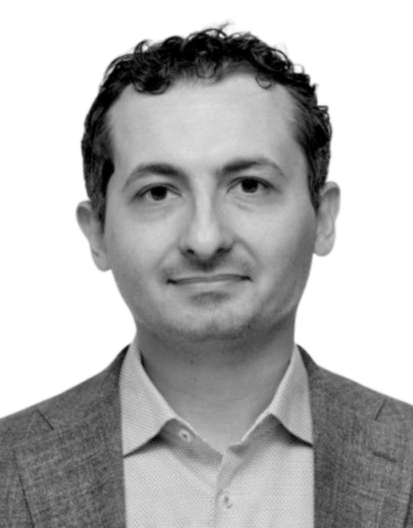} is the Co-founder and CTO of Kayhan Space. Kayhan Space is a software company focused on making spaceflight safer by delivering next-generation autonomous spaceflight capabilities. Araz has a master’s in Information Systems and has been a part of two successful startups. He has over 10 years of experience in technology and analytics working for startups, as well as Fortune 500 companies.
\end{biographywithpic}

\begin{biographywithpic}{Mykel J. Kochenderfer}{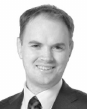}
is an Associate Professor of Aeronautics and Astronautics and Associate
Professor, by courtesy, of Computer Science at Stanford University. He is the
director of the Stanford Intelligent Systems Laboratory (SISL), conducting
research on advanced algorithms and analytical methods for the design of
robust decision-making systems. Prior to joining the faculty in 2013, he was
at MIT Lincoln Laboratory, where he worked on airspace modeling and aircraft
collision avoidance. He received his Ph.D. from the University of Edinburgh
in 2006, where he studied at the Institute of Perception, Action and Behaviour
in the School of Informatics. He received B.S. and M.S. degrees in computer
science from Stanford University in 2003.
\end{biographywithpic}
\endgroup

\end{document}